\documentclass{iau} 
\usepackage{graphicx}
\usepackage{natbib}
\def\aj{\textit{AJ}}                   
\def\araa{\textit{ARA\&A}}             
\def\apj{\textit{ApJ}}                 
\def\apss{\textit{Ap\&SS}}             
\def\apssp{\textit{Ap\&SS Proceedings}}    
\def\aap{\textit{A\&A}}                
\def\an{\textit{Astronomische Nachrichten}} 
\def\mnras{\textit{MNRAS}}             
\def\pasp{\textit{PASP}}               
\def\nat{\textit{Nature}}              
\title[What will eROSITA reveal among X-ray faint neutron stars?]
{What will eROSITA reveal among X-ray faint isolated neutron stars?}

\author[A.~M.~Pires]
{Adriana M.~Pires} 

\affiliation{
Leibniz-Institut f\"ur Astrophysik Potsdam (AIP), 
   An der Sternwarte 16, 14482, Potsdam, Germany \\ 
   email: {\tt apires@aip.de} 
}
\pubyear{2017}
\volume{337}  
\jname{Pulsar Astrophysics -­ The Next 50 Years}
\editors{P. Weltevrede, B.B.P. Perera, L. Levin Preston \& S. Sanidas, eds.}
\begin{document}
\maketitle
\begin{abstract}
Since the discovery of the first radio pulsar fifty years ago, the population of neutron stars in our Galaxy has grown to over 2,600. A handful of these sources, exclusively seen in X-rays, show properties that are not observed in normal pulsars. Despite their scarcity, they are key to understanding aspects of the neutron star phenomenology and evolution. The forthcoming all-sky survey of eROSITA will unveil the X-ray faint end of the neutron star population at unprecedented sensitivity; therefore, it has the unique potential to constrain evolutionary models and advance our understanding of the sources that are especially silent in the radio and $\gamma$-ray regimes. In this contribution I discuss the expected role of eROSITA, and the challenges it will face, at probing the galactic neutron star population. 
\keywords{surveys, stars: neutron, pulsars: general}
\end{abstract}
\firstsection 
\section{Historic overview: ROSAT}
Fifty years after the discovery of PSR~B1919+21 \citep{hew68}, most of the neutron stars in the Milky Way are still observed through their emission of pulsed radio waves. 
The ever increasing sensitivity and timing precision of pulsar surveys, most notably conducted at the Parkes and Arecibo radio telescopes, have greatly contributed to the discovery of over 2,600 neutron stars that are known to-date \citep{man05}. Besides, thanks to the Fermi `pulsar revolution' \citep{car14}, over 200 pulsars are now known to emit at high energies.
New detections have benefited from improved periodicity searching tecniques that can unveil transients, as well as sources with irregular pulse amplitude or long nulling fractions \citep[e.g.][]{lau06,kei09,kea10}. 
A ten-fold increase on the number of neutron stars will result from the operation of the Square Kilometre Array (SKA), which will potentially detect all active radio pulsars that are beamed towards the Earth \citep{kea15}.

Despite these succesful radio and $\gamma$-ray pulsar surveys, it is well known that the observed neutron star sample is biased by selection effects, and not representative of the true galactic population.
The Milky Way is expected to host over $10^8$ neutron stars \citep{die06}; averaged over its age, most of these compact objects will be old and, without a close companion, cannot be observed through radiative processes. 
The launch of ROSAT \citep{vog99}, a satellite of outstanding sensitivity at soft X-ray energies, raised a major expectation that a large ($\gtrsim10^3$) population of old isolated neutron stars (INSs) would be detected through Bondi-Hoyle accretion from the interstellar medium \citep[see][for a review]{tre00}. 

These optimistic expectations had to be critically reevaluated when only seven soft X-ray sources lacking evident counterparts in other wavelengths were identified in the ROSAT All-Sky Survey data \citep[e.g.][]{hab07}. Rather than old accretors, the seven sources turned out to be a local group of cooling neutron stars; their remarkable properties and elusiveness\footnote{Despite several searches for similar sources in ROSAT data \citep[e.g.][]{rut03,agu11}.} have earned them the nickname the `magnificent seven' (M7).
\begin{figure}[t]
\begin{center}
\includegraphics*[width=0.75\textwidth]{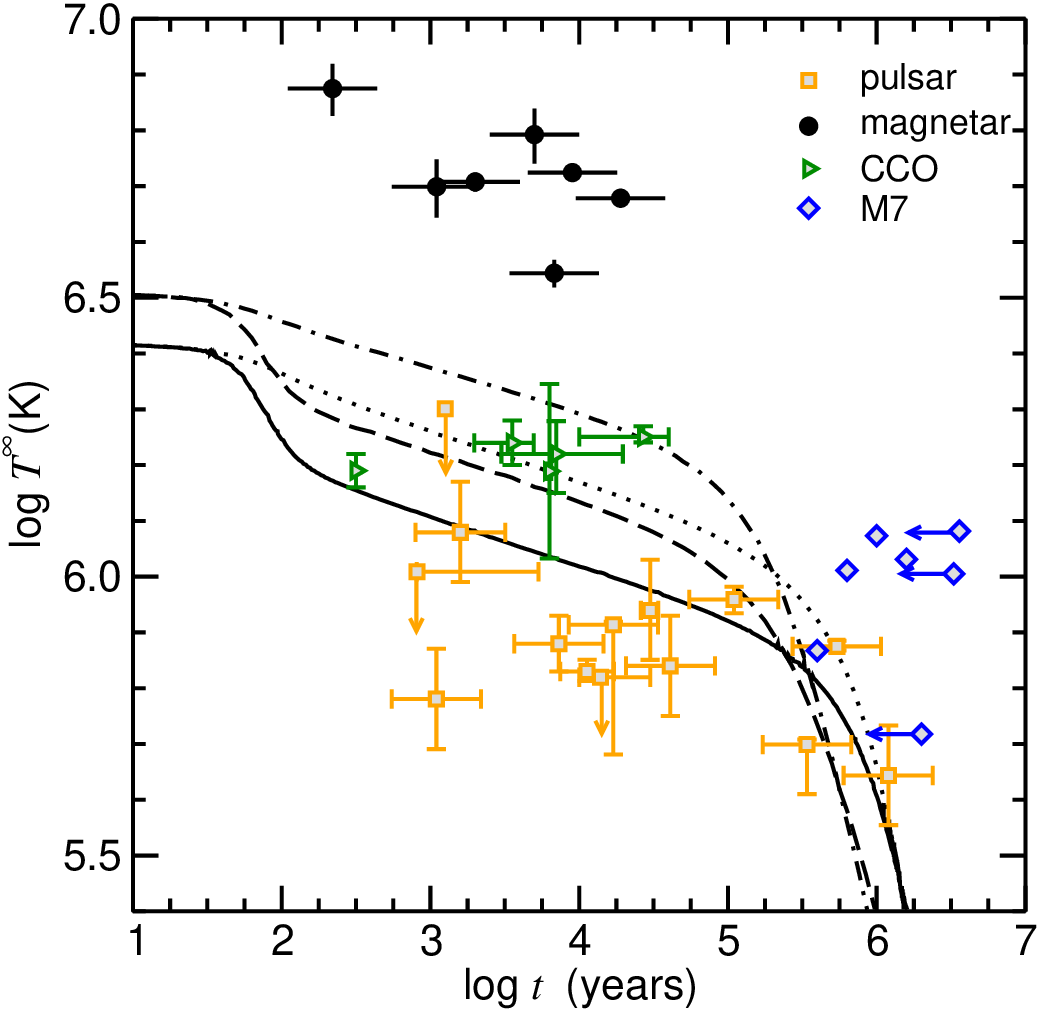}
\end{center}
\caption{Cooling-age diagram for INSs with thermal emission, in relation to the cooling scenario with no magnetic field decay (solid, dashed, dotted, and dotted-dahsed lines).
\label{fig_ccurve}}
\end{figure}
\section{Thermally emitting INSs and their evolution}
At present, around 50 INSs are known from their bright thermal X-ray emission and lack of magnetospheric radiation \citep[for an overview, see][]{mer11}. They are the sources known as magnetars (soft gamma repeaters and anomalous X-ray pulsars; see e.g.~\citealt{kas17}, for a review), the M7, and the central compact objects in supernova remnants (CCOs; \citealt{got13}). Despite their paucity, the peculiar groups are important to the understanding of evolutionary aspects that are unpredicted by theory and unobserved in the normal pulsar population.

Magnetars, a group comprising at present about 30 sources, are young ($\sim1-10$\,kyr) and energetic INSs powered by magnetic energy. Their phenomenology is complex and includes the emission of giant flares, dramatic spin down, bursts with the presence of bright spectral lines and quasi periodic oscillations, glitches, timing noise, multiwavelength variability, and pulse profile changes.

The X-ray emission of the M7 is believed to come from the still hot surface of a middle-aged ($0.1-1$\,Myr) neutron star. They rotate more slowly, have higher magnetic field intensities than the bulk of the rotation-powered pulsar population, and higher temperatures than expected for their ages (see cooling curve in Figure~\ref{fig_ccurve}). To a certain extent, such properties resemble those of magnetars. Magneto-thermal evolutionary models can account for their spectral and timing properties if they were born as magnetars and evolved through magnetic field decay \citep[e.g.][]{pon09,vig13}. 
\begin{figure}[t]
\begin{center}
\includegraphics*[width=0.495\textwidth]{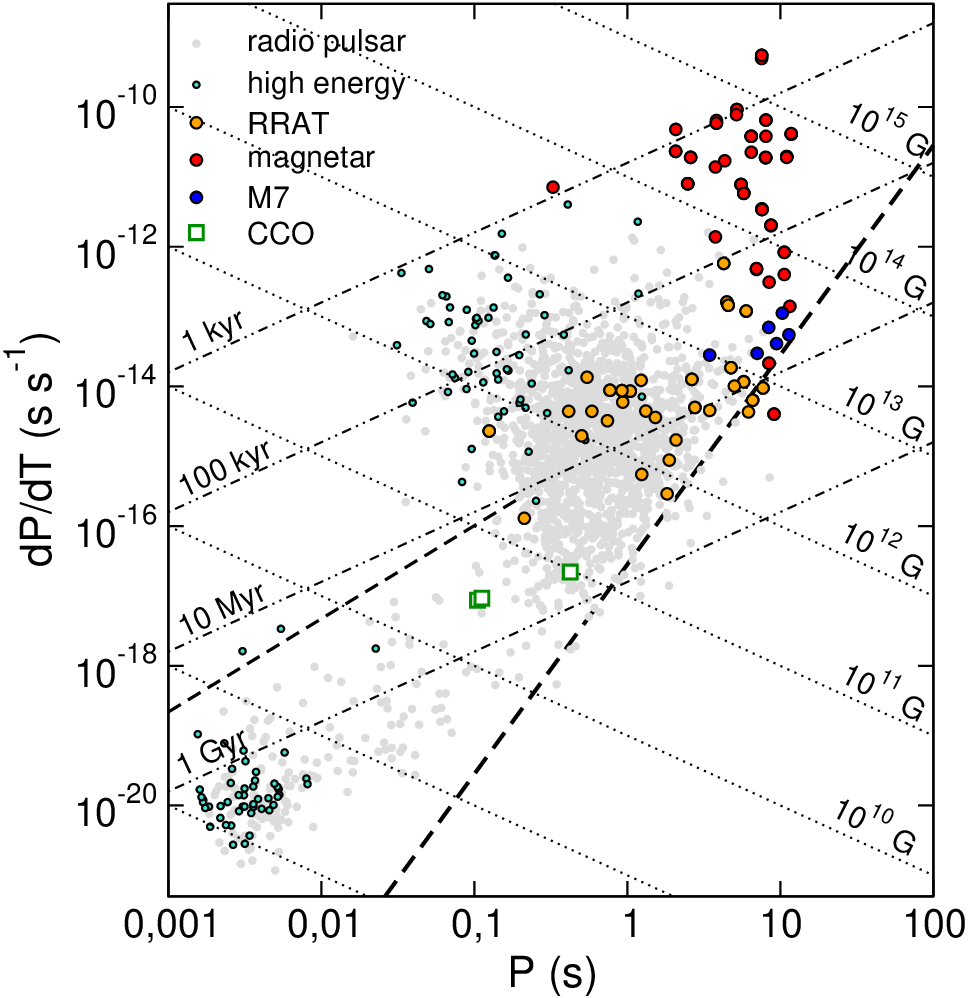}
\includegraphics*[width=0.495\textwidth]{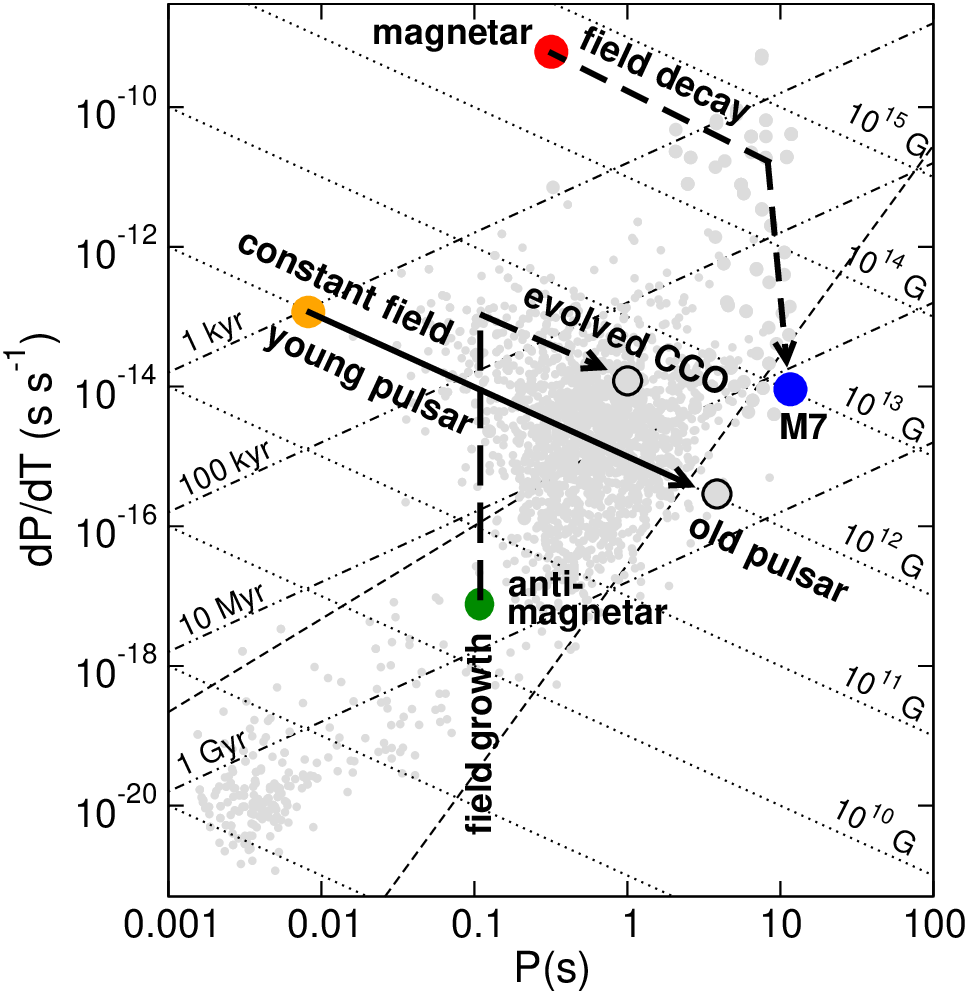}
\end{center}
\caption{\textit{Left.} $P-\dot{P}$ diagram of the galactic population of neutron stars, highlighting the position of peculiar neutron stars against the background of radio pulsars (see legend). \textit{Right.} Possible evolutionary scenarios, according to the magnetic field evolution.\label{fig_ppdot}}
\end{figure}

CCOs were first observed as point-like, radio-quiet X-ray sources located near the centre of supernova remnants. Of the dozen objects known at present, including candidates, at least three fit in the `anti-magnetar' scenario, in which these sources are young (few kyr) and weakly magnetised ($\sim10^{10}-10^{11}$\,G) neutron stars, experiencing very low spin down. There is growing observational evidence \citep[e.g.][]{bog14,luo15} that they might be recovering from an early episode of fallback accretion, which is responsible for burying the neutron star external magnetic field for several $\sim1-100$\,kyr \citep[e.g.][]{how11}. As a result, the neutron star shows in its early evolution the typical anti-magnetar behaviour. The pulsar is expected to join the rest of the population and evolve in the $P-\dot{P}$ diagram as the field reemerges, according to its original strength (Figure~\ref{fig_ppdot}). The episode might significantly alter the neutron star's cooling and perception as a thermal source due to the accreted light-element envelope \citep{yak04}. 
\section{eROSITA: forecast and harvesting}
The Extended R\"ontgen Survey with an Imaging Telescope Array \citep[eROSITA,][]{pre17} is the primary instrument on the forthcoming Spectrum-R\"ontgen-Gamma mission, which is expected to be launched in September 2018. In the four years following launch, eROSITA will survey the X-ray sky with unprecedented sensitivity, spectral and angular resolution, and is expected to yield a large sample of clusters of galaxies, active galactic nuclei, and Galactic compact objects, among other interesting case studies. 
The eROSITA All-Sky Survey (eRASS) represents therefore a timely opportunity for a better sampling of a considerable number of neutron stars that cannot be assessed by radio and $\gamma$-ray surveys. Eventually, the identification and investigation of eROSITA sources at faint X-ray fluxes will help us to test alternative neutron star evolutionary scenarios and constrain the rate of core-collapse supernovae in our Galaxy.

To estimate the number of isolated neutron stars to be detected in the eRASS through their thermal X-ray emission, we performed Monte Carlo simulations of a population synthesis model \citep{pir17}. 
Our study indicates an expected number of up to 100 thermally emitting neutron stars to be detected in the eRASS. 
Although optical follow-up will require very deep observations -- in particular, the identification of the faintest candidates will have to wait for the next generation of extremely large telescopes -- sources at intermediate fluxes can be selected for follow-up investigations using current facilities. Based on that, we anticipate $\sim$\,25 discoveries in the first years after the completion of the  all-sky survey.

The selection of newly proposed INS candidates among the myriad of eROSITA-detected sources and transient phenomena will be challenging: their identification and characterisation will require synergy with multiwavelength, large-scale, photometric and spectroscopic surveys that are operational in the near future. Moreover, state-of-the-art observing facilities, in particular 8-m-class optical telescopes, the XMM-Newton and Chandra Observatories, are crucial for dedicated follow-up campaigns to be conducted already in the immediate aftermath of the eROSITA survey.
\section*{Acknowledgements}
I thank Axel~Schwope, Christian~Motch, and Frank~Haberl for valuable discussions. 
This work made use of the ATNF pulsar catalogue. AMP is financially supported by the Deutsches Zentrum f\"ur Luft- und Raumfahrt (DLR) under grant 50 OR 1511.


\begin{thebibliography}{}
\bibitem[Ag\"ueros et~al.(2011)]{agu11}
Ag\"ueros, M.~A., Posselt, B., Anderson, S.~F., et~al. 2011, \aj, 141, 176

\bibitem[Bogdanov(2014)]{bog14}
Bogdanov, S. 2014, \apj, 790, 94

\bibitem[Caraveo(2014)]{car14}
Caraveo, P.~A. 2014, \araa, 52, 211

\bibitem[Diehl et~al.(2006)]{die06}
Diehl, R., Halloin, H., Kretschmer, K., et~al. 2006, \nat, 439, 45

\bibitem[Gotthelf et~al.(2013)]{got13}
Gotthelf, E.~V., Halpern, J.~P., \& Alford, J. 2013, \apj, 765, 58

\bibitem[Haberl(2007)]{hab07}
Haberl, F. 2007, \apss, 308, 181

\bibitem[Hewish et~al.(1968)]{hew68}
Hewish, A., Bell, S.~J., Pilkington, J.~D., Scott, P.~F., \& Collins, R.~A. 1968, \nat, 217, 709

\bibitem[Ho(2011)]{how11}
Ho, W.~C.~G. 2011, \mnras, 414, 2567

\bibitem[Kaspi \& Beloborodov(2017)]{kas17}
Kaspi, V.~M. \& Beloborodov, A. 2017, \araa, in press (ArXiv e-prints 1703.00068)

\bibitem[Keane et~al.(2010)]{kea10}
Keane, E.~F., Ludovici, D.~A., Eatough, R.~P., et~al. 2010, \mnras, 401, 1057

\bibitem[Keane et~al.(2015)]{kea15}
Keane, E., Bhattacharyya, B., Kramer, M., et~al. 2015, in Advancing Astrophysics 
 with the Square Kilometre Array, id.~40

\bibitem[Keith et~al.(2009)]{kei09}
Keith, M.~J., Eatough, R.~P., Lyne, A.~G., et~al. 2009, \mnras, 395, 837

\bibitem[Luo et~al.(2015)]{luo15}
Luo, J., Ng, C.-Y., Ho, W.~C.~G., et~al. 2015, \apj, 808, 130

\bibitem[Manchester et~al.(2005)]{man05}
Manchester, R.~N., Hobbs, G.~B., Teoh, A., \& Hobbs, M. 2005, \aj, 129, 1993

\bibitem[McLaughlin et~al.(2006)]{lau06}
McLaughlin, M.~A., Lyne, A.~G., Lorimer, D.~R., et~al. 2006, \nat, 439, 817

\bibitem[Mereghetti(2011)]{mer11}
Mereghetti, S. 2011, \apssp, 21, 345

\bibitem[Pires et al.(2017)]{pir17}
Pires, A.~M., Schwope, A.~D., \& Motch, C. 2017, \an, 338, 213

\bibitem[Pons et al.(2009)]{pon09}
Pons, J.~A., Miralles, J.~A. and Geppert, U. 2009, \aap, 496, 207

\bibitem[Predehl(2017)]{pre17}
Predehl, P. 2017, \an, 338, 159

\bibitem[Rutledge et~al.(2003)]{rut03}
Rutledge, R.~E., Fox, D.~W., Bogosavljevic, M., \& Mahabal, A. 2003, \apj, 598, 458

\bibitem[Treves et~al.(2000)]{tre00}
Treves, A., Turolla, R., Zane, S., \& Colpi, M. 2000, \pasp, 112, 297

\bibitem[Vigan\`o et~al.(2013)]{vig13}
Vigan\`o, D., Rea, N., Pons, J.~A., et~al. 2013, \mnras, 434, 123

\bibitem[Voges et~al.(1999)]{vog99}
Voges, W., Aschenbach, B., Boller, T., et~al. 1999, \aap, 349, 389

\bibitem[Yakovlev \& Pethick(2004)]{yak04}
Yakovlev, D.~G. \& Pethick, C.~J. 2004, \araa, 42, 169
\end{thebibliography}
\end{document}